\documentclass[aps,prb,twocolumn,superscriptaddress]{revtex4-1}

\usepackage{graphicx}
\usepackage{feynmp}
\usepackage{epsfig}
\usepackage[latin1]{inputenc}
\usepackage{amsmath}
\usepackage{amsfonts}
\usepackage{amssymb}
\usepackage{amsmath}
\usepackage{amssymb}
\usepackage{color}
\usepackage{multirow,bigdelim}
\usepackage{enumerate}
\usepackage{amsthm}
\usepackage{amsbsy}
\usepackage{wasysym}
\usepackage{mathrsfs}
\usepackage{subfigure}

\newcommand{\be}{\begin{equation} }
\newcommand{\ee}{\end{equation} }
\newcommand{\ba}{\begin{eqnarray} }
\newcommand{\ea}{\end{eqnarray} }

\begin{document}

\title{On equilibration and coarsening in the quantum $O(N)$ model at infinite $N$}

\author{Anushya Chandran}
\affiliation{Department of Physics, Princeton University, Princeton, New Jersey 08544, USA}
\author{Arun Nanduri}
\affiliation{Department of Physics, Princeton University, Princeton, New Jersey 08544, USA}
\author{S. S. Gubser}
\affiliation{Department of Physics, Princeton University, Princeton, New Jersey 08544, USA}
\author{S. L. Sondhi}
\affiliation{Department of Physics, Princeton University, Princeton, New Jersey 08544, USA}
\date{\today}

\begin{abstract}
The quantum $O(N)$ model in the infinite $N$ limit is a paradigm for symmetry-breaking. Qualitatively, its phase diagram is an excellent guide to the equilibrium physics for more realistic values of $N$ in varying spatial dimensions ($d>1$). Here we investigate the physics of this model out of equilibrium, specifically its response to global quenches starting in the disordered phase. If the model were to exhibit equilibration, the late time state could be inferred from the finite temperature phase diagram. In the infinite $N$ limit, we show that not only does the model not lead to equilibration on account of an infinite number of conserved quantities, it also does \emph{not} relax to a generalized Gibbs ensemble consistent with these conserved quantities. Nevertheless, we \emph{still} find that the late time states following quenches bear strong signatures of the equilibrium phase diagram. Notably, we find that the model exhibits coarsening to a non-equilibrium critical state only in dimensions $d>2$, that is, if the equilibrium phase diagram contains an ordered phase at non-zero temperatures.

\end{abstract}
\maketitle
\section{Introduction}
\begin{figure*}[htb]
\begin{minipage}[b]{0.45\linewidth}
\centering
\includegraphics[width=\textwidth]{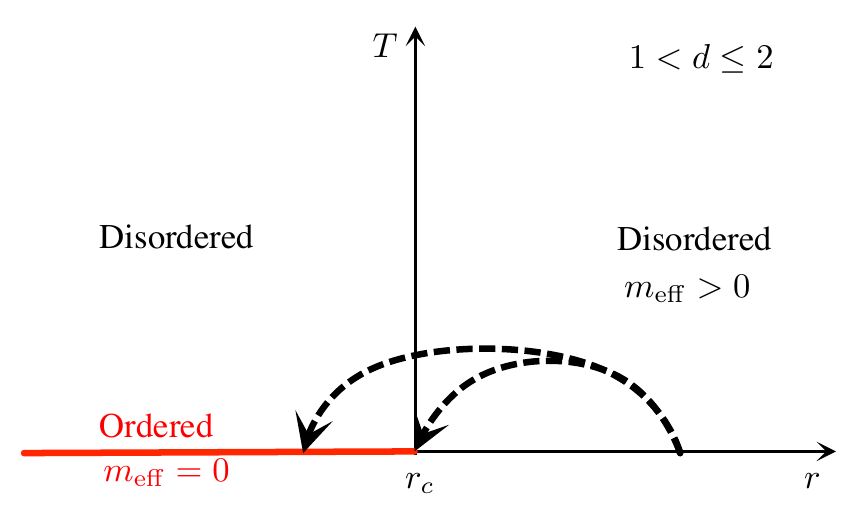}
\caption{Topology of the equilibrium phase diagram of the $O(N)$ model in
the infinite $N$ limit in spatial dimensions $1<d\leq 2$. The dashed lines indicate the different quenches that we study. }
\label{fig:phased_1}
\end{minipage}
\hspace{0.5cm}
\begin{minipage}[b]{0.45\linewidth}
\centering
\includegraphics[width=\textwidth]{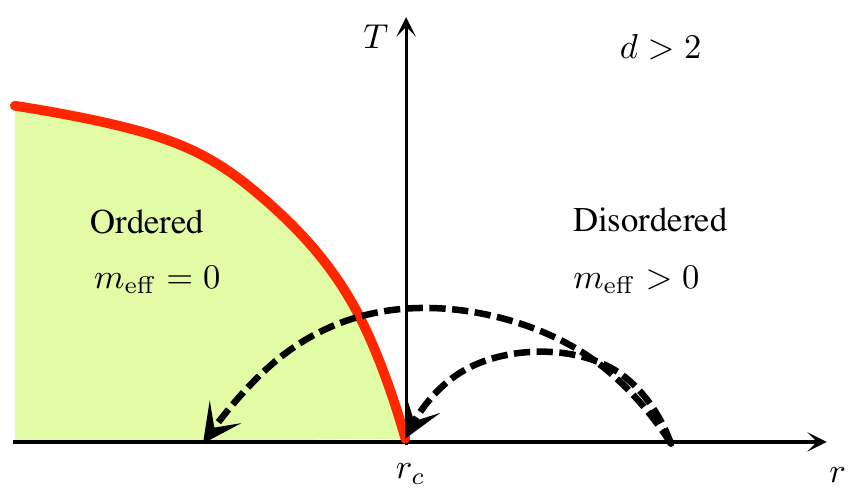}
\caption{Topology of the equilibrium phase diagram of the $O(N)$ model in
the infinite $N$ limit in spatial dimensions $d>2$. The dashed lines indicate the different quenches that we study.}
\label{fig:phased_2}
\end{minipage}
\end{figure*}

The study of the out of equilibrium dynamics of closed quantum systems has
intensified greatly in recent years inspired in considerable measure by
advances in the experimental study of cold atomic systems \cite{Greiner:2002aa, Greiner:2002zr, Sadler:2006aa,Trotzky:2012uq, Sabbatini:2012vn} (see Ref.~\onlinecite{Bloch:2008ly, Dziarmaga:2010aa, Polkovnikov:2011ys} for recent reviews). One important theme
in this work is the presence and nature of equilibration starting from a
non-equilibrium state and its relationship to integrability. An interesting
stream of work has postulated and examined the notion of a generalized Gibbs
ensemble (GGE) in which an integrable system relaxes to a maximum entropy
state consistent with {\it all} of its constants of motion \cite{Rigol:2007cr, Kinoshita:2006ve}. A second important
theme is the interplay between non-equilibrium dynamics and phase structure
which goes under the moniker of the Kibble-Zurek problem\cite{Kibble:1976aa, Zurek:1996aa}, wherein the system
is driven out of equilibrium by non-adiabatic changes of parameters, typically
between values in different phases or between a phase and a critical point, see e.g. Refs.~\onlinecite{Polkovnikov:2005zr, Chandran:2012qa, De-Grandi:2011lq}.

A large fraction of this work has concerned itself, for natural reasons of
tractability, to systems in spatial dimension $d=1$ where analytic \cite{Calabrese:2006aa, Cazalilla:2006aa, Cramer:2008ve, De-Grandi:2010aa, Mitra:2012qf, Calabrese:2011ys} and computational power \cite{Rigol:2009bh, Kolodrubetz:2012kx}
can be readily brought to bear. This restriction to $d=1$ however does not allow the study of the impact of dimensionality, known to be important
for equilibrium behavior. In this paper, we study a model where one
{\it can} move between dimensions while retaining tractability---the quantum $O(N)$ vector model in the infinite $N$ limit (see Ref.~\onlinecite{Moshe:2003aa} for a comprehensive introduction). This model has much to
commend it. It yields an equilibrium phase diagram in the Gibbs ensemble
in various dimensions whose topology is correct for $N \ge 3$. It also
yields critical exponents which incorporate corrections beyond
Gaussian critical behavior in $d < 3$ and correctly locates the lower and upper
critical dimension at $d=1$ and $d=3$ respectively.

The above results {\it assume} that the Gibbs ensemble is reached. In this
paper we study the late time states of the model starting out of
equilibrium and ask to what extent the equilibrium phase diagram is
a guide to the late time behavior. We note that the model has certainly
been the object of prior study, in the first instance from a cosmology
inspired interest in non-equilibrium field theory \cite{Boyanovsky:1996kx, Cooper:1997mi, Bettencourt:1998pi,Boyanovsky:1999uq} and more recently from
the condensed matter/statistical mechanical viewpoint \cite{Sotiriadis:2010ys, Das:2012kx, Sciolla:2012ys}. We build on this
work but find that there are still new things to say on this problem, largely
due to asking some fresh questions from the condensed matter/statistical mechanical perspective.

Our results organize themselves naturally into the two themes we noted at
the outset. First, we revisit the question of equilibration in the infinite $N$
vector model. It has been observed previously that the model has a large
number of conserved quantities that will prevent a relaxation to a Gibbs
ensemble. We show further that the model also does {\it not} relax to the GGE
that builds in these conserved quantities. This behavior is in contrast to that of the non-interacting, purely Gaussian, vector model which does relax to a
GGE, though now with a larger number of conserved quantities \cite{Calabrese:2007aa,Barthel:2008zr}. Second, we
examine spatially homogenous quenches starting from ground states in the disordered phase to the
critical coupling and into the couplings in the ordered phase. For the latter,
we show that the late time state exhibits coarsening in the sense of a diverging equal time correlation length in $d > 2$ for sufficiently gentle quenches.
Our numerics further suggest that the system coarsens towards a non-equilibrium critical state as $t\rightarrow \infty$.
For larger quenches in $d>2$ or for any quenches in $d \le 2$, we find that no coarsening
is possible. Instead, the late time state is disordered. For quenches to the critical point, we find no coarsening for
all $d$ but coarsening in the scaling limit for $d \ge 3$ consistent with
the lack of scattering at the Gaussian fixed point and previous results on
the Gaussian theory \cite{Calabrese:2007aa}. Interestingly, the results on quenches are qualitatively
what one would predict assuming equilibration following an injection of
energy density and yet they hold for a system that does not equilibrate.

In the following, we document these claims. We begin with a review of
the infinite $N$ vector model and the equations that must be solved to determine
its dynamics in Section~\ref{Sec:TheModel}. In Section~\ref{Sec:Ergodicity}, we discuss the issue of ergodicity
or equilibration. In Section~\ref{Sec:ToOrderedPhase}, we describe our results on global quenches into
the ordered phase. In Section~\ref{Sec:ToCriticalPoint}, we discuss global quenches to the critical point and
show that the stability of the Gaussian fixed point for $d \ge 4$ when it comes
to equilibrium behavior also shows up in the behavior of non-equilibrium
quenches provided one takes a scaling limit discussed in Ref~\onlinecite{Chandran:2012qa}.
We finish with some concluding remarks in Section~\ref{Sec:Conclusion} and relegate some
technical material to an appendix.

\section{The model}
\label{Sec:TheModel}

The Hamiltonian of the quantum $O(N)$ model in $d$ spatial dimensions is
\begin{align}
\label{Eq:O(N)H}
H= \frac{1}{2}\int d^d x \left(  |\vec{\Pi}|^2 + |\vec{\nabla} \vec{\Phi}|^2 + r  | \vec{\Phi}|^2 + \frac{\lambda}{2N }  | \vec{\Phi}|^4\right),
\end{align}
where $\vec{\Phi}$ and $\vec{\Pi}$ are canonically conjugate $N$-component fields,
\begin{align}
[\Phi_i(\vec{x}), \Pi_j(\vec{x}') ] = i \delta^d(\vec{x}-\vec{x}') \delta_{ij}.
\end{align}
In the limit $N\rightarrow \infty$, the equilibrium physics is soluble. In the disordered phase, $\langle \vec{\Phi}\rangle = 0$, the ground state of $H$ is well-approximated by the ground state of a free field theory with mass $m_{\rm eff}$ that is determined self-consistently:
\begin{align}
m_{\rm eff}^2 &= r + \lambda \left\langle \frac{|\vec{\Phi}|^2}{N} \right\rangle \\
\left\langle \frac{|\vec{\Phi}|^2}{N} \right\rangle &= \int^\Lambda \frac{d^dk}{(2\pi)^d} \frac{1}{2\sqrt{|\vec{k}|^2 + m_{\rm eff}^2}}, \label{Eq:meff2Eq}
\end{align}
where $\Lambda$ is the cutoff. The effective mass $m_{\rm eff}$ is proportional to the single-particle gap and controls the phase diagram (Figs.~\ref{fig:phased_1}, \ref{fig:phased_2}). In the disordered phase, $m_{\rm eff}>0$, while in the ordered phase, when suitably re-defined, $m_{\rm eff}$ is the mass of the Goldstone bosons and equals zero. We describe the phase diagram in greater detail in Sec.~\ref{Sec:ToOrderedPhase}. The solubility in equilibrium can be traced to two related sources: 1) The ground state is a (Gaussian) symmetric product state over component indices, and 2) the expectation value:
\begin{align}
\left\langle \frac{|\vec{\Phi}|^2}{N} \Phi_i \right\rangle = \left\langle \frac{|\vec{\Phi}|^2}{N} \right\rangle \langle\Phi_i \rangle + O\left(\frac{1}{N}\right) \label{Eq:factorizationEq}
\end{align}
factorizes to leading order in $1/N$. These two features of the infinite $N$ model in equilibrium make the dynamical problem tractable as well.

Now consider preparing the system at $t=0$ in a product state symmetric in the component indices\footnote{More generally any state, such as the ground
state for some value of the bare parameters, which is well approximated by
such a product state. The dynamics is also tractable for the class of density
matrices that have product structure in the component index}:
\begin{align}
|\psi(0)\rangle = \prod_{i=1}^N |\beta\rangle_i. \label{Eq:InitState}
\end{align}
The wavefunction of each component, $|\beta\rangle$, can be freely chosen. In particular, $|\psi(0)\rangle$ need not be Gaussian or be a ground state of the infinite $N$ model anywhere in the disordered phase. In general, it does not satisfy Wick's theorem. The evolution of the state $|\psi(0)\rangle$ for $t>0$ is generated by $H$. The Heisenberg equations of motion are:
\begin{align}
\frac{d\Phi_i}{dt} &= \Pi_i \nonumber\\
\frac{d\Pi_i}{dt} &= \nabla^2 \Phi_i - r \Phi_i - \frac{\lambda}{N} |\vec{\Phi}|^2 \Phi_i.\label{Eq:Heisenberg}
 \end{align}
For the rest of this article, we work in the Heisenberg picture. Remarkably, the factorization in equilibrium in Eq.~\eqref{Eq:factorizationEq} holds out-of-equilibrium as well:
\begin{align}
\left\langle \frac{|\vec{\Phi}(t)|^2}{N} \Phi_i(t) \right\rangle = \left\langle \frac{|\vec{\Phi}(t)|^2}{N} \right \rangle \langle \Phi_i(t)\rangle + O\left(\frac{1}{N}\right).
\end{align}
As in equilibrium, this factorization leads to an effective mass in a free field theory that is determined self-consistently. However, the effective mass is now time-dependent. More formally, consider the Hamiltonian for a free field theory with the time-dependent mass $m_{\rm eff}(t)$ determined self-consistently at each $t$:
\begin{align}
H_{\rm eff}(t) &= \frac{1}{2}\int^\Lambda \frac{d^d k}{(2\pi)^d}\,  \left[|\vec{\Pi}_{\vec{k}}|^2 + (\,|\vec{k}|^2 + m_{\rm eff}^2(t) \,) | \vec{\Phi}_{\vec{k}}|^2\right] \label{Eq:HeffDef} \\
m_{\rm eff}^2(t)& \equiv r + \lambda \left\langle \frac{|\vec{\Phi}(t)|^2}{N} \right\rangle. \label{Eq:meffDef}
\end{align}
Above, $[\Phi_i(\vec{k}), \Pi_j(\vec{k}') ] = i(2\pi)^d \delta^d(\vec{k}-\vec{k}') \delta_{ij}$. We evolve the state $|\psi(0)\rangle$ with $H_{\rm eff}$. We also evolve $|\psi(0)\rangle$ with $H$ in the limit $N\rightarrow \infty$. The formal statement is that all correlation functions involving a finite number of components are identical in the two cases at any fixed $t$. Thus, to determine observables and correlation functions, we need only solve for the dynamics in a free field theory with mass $m_{\rm eff}(t)$. As in equilibrium, the infinite $N$ model out-of-equilibrium goes beyond the free field theory through the single self-consistency condition on $m_{\rm eff}(t)$ (Eq.~\eqref{Eq:meffDef}). Henceforth, we suppress the component index when an expectation value is independent of it. Specifically, we replace $ \langle |\vec{\Phi}(t)|^2\rangle /N$ by the expectation $\langle \Phi^2(t) \rangle$ of a single component. When not indicated, all operators in a correlation function are assumed to have the same component label.

A convenient way to determine correlation functions from $H_{\rm eff}(t)$ is to expand $\Phi_{\vec{k}}(t)$ and $\Pi_{\vec{k}}(t)$ in a fixed basis at $t=0$:
\begin{align}
&\Phi_{\vec{k}}(t) = \frac{f_{\vec{k}}(t)}{\sqrt{2}} \,a_{\vec{k}} + \frac{f^*_{-\vec{k}}(t)}{\sqrt{2}} \,a_{-\vec{k}}^\dagger \nonumber\\
&\Pi_{\vec{k}}(t) = \frac{d \Phi^\dagger_{\vec{k}}(t)}{dt}.\label{Eq:Fixedbasis}
\end{align}
Above, the $(a_{\vec{k}},a^\dagger_{\vec{k}})$ and $(a_{-\vec{k}},a^\dagger_{-\vec{k}})$ are two independent sets of fixed ladder operators with the usual commutation relations. The $f_{\vec{k}}(t)$ are complex-valued coefficients known as mode functions. The commutation relations stipulate that:
\begin{align}
f_{\vec{k}}(t) &= f_{-\vec{k}}(t) \nonumber \\
\textrm{Im}[f_{\vec{k}}(t) \, \dot{f}_{\vec{k}}^*(t)] &= 1. \label{Eq:Wronskian}
\end{align}
Given an initial state, we may choose any fixed basis to decompose the field operators in. This is a coordinate choice; in this article, we pick the particular fixed basis in which:
\begin{align}
\langle a_{\vec{k}} a_{\vec{k}'} \rangle &=0, \qquad f_{\vec{k}}(0) = f_{\vec{k}}^*(0). \label{Eq:NewGauge}
\end{align}
It is then easy to see that $f_{\vec{k}}(0)$ and $|\dot{f}_{\vec{k}}(0)|$ fix all the coefficients in Eq.~\eqref{Eq:Fixedbasis} at $t=0$. For example, Eq.~\eqref{Eq:Wronskian} determines the phase of $\dot{f}_{\vec{k}}(0)$ etc. Both functions follow from the two-point functions in the initial state:
\begin{align}
\langle \Phi_{\vec{k}}(0)  \Phi^\dagger_{\vec{k}'}(0) \rangle &= \frac{f_{\vec{k}}(0)^2 \mathcal{N}_{\vec{k}}}{2}  \, (2\pi)^d \delta^d(\vec{k}-\vec{k}') \nonumber  \\
\langle \Pi_{\vec{k}}(0)  \Pi^\dagger_{\vec{k}'}(0) \rangle&= \frac{|\dot{f}_{\vec{k}}(0)|^2  \mathcal{N}_{\vec{k}} }{2}\,(2\pi)^d\delta^d(\vec{k}-\vec{k}') \nonumber \\
 Re[\langle \Phi_{\vec{k}}(0) \Pi_{\vec{k}'}(0) \rangle] &=  Re[\dot{f}_{\vec{k}}(0)] f_{\vec{k}}(0)  \mathcal{N}_{\vec{k}} \, (2\pi)^d\delta^d(\vec{k}-\vec{k}'). \label{Eq:cnumberst0}
\end{align}
if we substitute for the phase of $\dot{f}_{\vec{k}}(0)$ from Eq.~\eqref{Eq:Wronskian}. $\mathcal{N}_{\vec{k}}$ above is related to the sum of the occupations of the $\pm \vec{k}$ modes in the initial state:
\begin{align*}
\mathcal{N}_{\vec{k}} &\equiv 1+n_{\vec{k}} + n_{-\vec{k}} \textrm{ , } \langle a_{\vec{k}}^\dagger a_{\vec{k}'}\rangle = n_{\vec{k}} (2\pi)^d \delta^d (\vec{k}-\vec{k}').
\end{align*}
Finally, from $H_{\rm eff}(t)$, the definition of $m_{\rm eff}^2$ in Eq.~\eqref{Eq:meffDef}, and Eq.~\eqref{Eq:Fixedbasis} , we obtain the equations of motion for the mode functions:
\begin{align}
&\left(\frac{d^2}{dt^2} + |\vec{k}|^2 + m_{\rm eff}^2(t)\right)f_{\vec{k}}(t) =0 \label{Eq:ModeFunction}\\
m_{\rm eff}^2(t) &=r +\frac{ \lambda}{2} \int \frac{d^dk}{(2\pi)^d} |f_{\vec{k}}|^2 \mathcal{N}_{\vec{k}}.   \label{Eq:meff2inf}
\end{align}
Given the correlation functions in Eq.~\eqref{Eq:cnumberst0} and the condition in Eq.~\eqref{Eq:Wronskian}, we can propagate this system of coupled second-order differential equations forward in time to solve for $f_{\vec{k}}(t)$. We may then compute any correlation function of interest by means of Eq.~\eqref{Eq:Fixedbasis} and the solutions for the mode functions.

\section{Ergodicity}
\label{Sec:Ergodicity}
To determine the nature of the late time state, we need to identify all the quantities that are conserved in the dynamics (Eq.~(\ref{Eq:ModeFunction},\ref{Eq:meff2inf})). First, as the evolution is generated by a Hamiltonian, the total energy is conserved. This would be true for any isolated quantum system. Additionally at infinite $N$, the operators $L_{\vec{k}}^z$ defined below also commute with $H$ in the class of states in Eq.~\eqref{Eq:InitState}.
\begin{align}
L^z_{\vec{k}}\equiv \frac{\Phi_{\vec{k}}^\dagger \Pi_{\vec{k}}^\dagger - \Phi_{\vec{k}} \Pi_{\vec{k}}}{i}.\label{Eq:QuantumLzDef}
 \end{align}
Thus,
 \begin{align}
 \frac{d\langle L^z_{\vec{k}} \rangle }{dt}=0.\label{Eq:AngMomConserved}
 \end{align}
We use $\vec{k}>0$ to denote half the momenta; each $\vec{k}$ in the set labels the pair $(\vec{k},-\vec{k})$. The conserved quantity $\langle L^z_{\vec{k}}\rangle $ has two interpretations. One is the angular momentum of the 2d harmonic oscillator at $\vec{k}>0$ in $H_{\rm eff}(t)$. The second is in the fixed basis in which $\langle L^z_{\vec{k}} \rangle=n_{\vec{k}} - n_{-\vec{k}}$. This difference is conserved as the scattering processes that lead to exchange of momenta between pairs of bosons is suppressed at infinite $N$.

There are therefore half as many conserved quantities as the number of degrees of freedom. This is reminiscent of an integrable system in which the number of conserved quantities equals the number of degrees of freedom. For example, the free field theory ($\lambda=0$) is integrable and has two conserved quantities for each $(\vec{k},-\vec{k})$ pair: the energy in the 2d oscillator and the angular momentum $L^z_{\vec{k}}$. In such systems, the correlations in the late time state are conjectured to be reproduced by the generalized Gibbs ensemble (GGE) \cite{Rigol:2007cr,Dziarmaga:2010aa} :
\begin{align*}
\rho_{GGE} = \exp(- \sum_{i=1}^M \mu_{i} O_i),
\end{align*}
where the $O_i$ commute with the Hamiltonian and the $\mu_i$ are fixed by $\langle O_i \rangle$ in the initial state. This conjecture has been checked for the two-point correlation functions in the free field theory (if Wick's theorem is satisfied in the initial state, higher point correlation functions would also be correctly reproduced).

The collection of conserved quantities in the $O(N)$ model at infinite $N$ suggests that the late time state should thermalize to the following GGE:
\begin{align}
\rho_{GGE} = \exp\left(-\beta H - \int_{\vec{k}>0} \mu_{\vec{k}} L^z_{\vec{k}} \right). \label{Eq:GGEON}
\end{align}
We argue in two (possibly related) ways below that the dynamics \emph{fails} to relax to this GGE.

First, note that Eq.~\eqref{Eq:meff2inf} depends on a c-number associated with the initial state. A reasonable expectation would be that correlation functions at late times also depend on this c-number. However, correlation functions derived from the GGE do not. This suggests that the GGE is not the correct description of the late-time dynamics. In Appendix~\ref{App:WhynoGGE}, we sharpen this intuition by showing that the structure factor at late times depends on $\mathcal{N}_{\vec{k}}$. It follows that two-point function in real-space is also a function of the $\mathcal{N}_{\vec{k}}$, thus proving the claim that the system does not relax to the GGE for all local observables. We note that Ref.~\onlinecite{Bettencourt:1998pi} arrived at the same conclusion by constructing a particular non-linear combination of equal-time correlation functions that were independent of time and involved $\mathcal{N}_{\vec{k}}$.

The second argument hinges on the numerical observation that $m_{\rm eff}^2(t)$ generically approaches a non-negative constant, $m_f^2$, as $t\rightarrow \infty$. This implies that the different $\vec{k}$ modes decouple and the theory is effectively free at late times. Thus, an extensive set of conserved quantities (equal to the number in the GGE) \emph{emerges} at late times:
\begin{align}
E_{\vec{k}}= |\vec{\Pi}_{\vec{k}}|^2 + (\,|\vec{k}|^2 + m_f^2 \,) | \vec{\Phi}_{\vec{k}}|^2 \label{Eq:NewConserved}
\end{align}
and the system appears integrable. These extra conserved quantities are independent of $\langle L^z_{\vec{k}}\rangle$: they are functions of c-numbers like $\mathcal{N}_{\vec{k}}$ in the initial state. Thus, the late-time state has more memory than the GGE about the initial conditions and the dynamics does not relax to the GGE in Eq.~\eqref{Eq:GGEON}.

This leaves open the possibility that there is a more restrictive GGE that
includes some unknown additional conserved quantities that {\it will}
describe the late time states. While we cannot rule this out, we (and
previous workers on this problem) have not been able to find such quantities.

\section{Quenches to the ordered phase}
\label{Sec:ToOrderedPhase}
\subsubsection{Problem, methods and background}

We now turn to the late time behavior of protocols starting from ground states in the disordered phase wherein the coupling is changed to a value in which the ground state is ordered (see Figs.~\ref{fig:phased_1} and \ref{fig:phased_2}).
For specificity and because our interest here is primarily in the late time behavior, we will take these protocols to
be sudden quenches, i.e.~the coupling will be changed instantly to its new value.  Our conclusions will generalize
{\it mutatis mutandis} to more elaborate protocols such as those of interest in the Kibble-Zurek problem. Right after the quench, the system is in an excited state for the new Hamiltonian, i.e.~one with a non-zero energy density. For a system which thermalizes, the late time state can be read off from the equilibrium phase diagram by converting the energy density into an equivalent temperature.
Interestingly, despite the lack of relaxation in the quantum $O(N)$ model at infinite $N$, we will find that the phase diagram
is {\it still} a good heuristic for the late time behavior when the system is quenched, in a sense we will shortly make precise.

But first, let us describe the setup of the problem more precisely. Recall that the mode functions are constrained by Eqs.~\eqref{Eq:Wronskian} and \eqref{Eq:NewGauge}. Polar coordinates make these constraints more transparent. Define $f_{\vec{k}}\equiv |f_{\vec{k}}| \exp(i \theta_{\vec{k}})$. Eq.~\eqref{Eq:Wronskian} then provides a relation between the amplitude and the phase of the mode function:
\begin{align}
\dot{\theta}_{\vec{k}} = \frac{-1}{|f_{\vec{k}}|^2}. \label{Eq:PhaseFromAmp}
\end{align}
Together with the initial condition $\theta_{\vec{k}}(0)=0$, this allows us to reconstruct the mode function at all $t$ from $|f_{\vec{k}}(t)|$ alone. The equation of motion for $|f_{\vec{k}}(t)|$ follows from Eq.~(\ref{Eq:ModeFunction},\ref{Eq:meff2inf}):
\begin{align}
&\left(\frac{d^2}{dt^2} + |\vec{k}|^2 + m_{\rm eff}^2(t)\right)|f_{\vec{k}}| - \frac{1}{|f_{\vec{k}}|^3} =0 \label{Eq:RhoEq}\\
m_{\rm eff}^2(t) &=r +\frac{ \lambda}{2} \int \frac{d^dk}{(2\pi)^d} |f_{\vec{k}}|^2 \, \mathcal{N}_{\vec{k}} \label{Eq:Meff2Rho}.
\end{align}
It suffices to solve these equations subject to the initial conditions discussed below to determine $f_{\vec{k}}(t)$.

We pick the initial state of the system for $t<0$ to be the ground state in the disordered phase. That is, we prepare the system in the ground state of $H$ (Eq.~\eqref{Eq:O(N)H}) with bare coupling $r=r_0$ greater than the critical coupling $r_c$. This state is the ground state of a free field theory with the effective mass $m_{0}$.  The relation between $m_{0}$ and $r_0$ is given by Eq.~\eqref{Eq:meff2Eq}. In this state:
\begin{align}
\mathcal{N}_{\vec{k}}&=1,\quad \Omega_{k0} = \sqrt{|\vec{k}|^2 + m_{0}^2} \nonumber\\
 f_{\vec{k}} (t<0) &=  \frac{1}{\sqrt{\Omega_{k0}}}, \quad \dot{f}_{\vec{k}} (t<0) = -i \sqrt{\Omega_{k0}}. \label{Eq:InitConds}
\end{align}
$\Omega_{k0}$ above is the frequency of the harmonic oscillator at $\vec{k}$. At $t=0$, we suddenly quench to the ordered phase ($r<r_c$). We note that in terms of our discussion earlier, the conserved angular momenta are all zero so that the associated chemical potentials $\mu_{\vec{k}}$ in the GGE are all also zero.

For the most part we will rely on numerical solutions of the dynamical equations. Specifically, we numerically solve Eqs.~(\ref{Eq:RhoEq},\ref{Eq:Meff2Rho}) subject to the initial conditions in Eq.~\eqref{Eq:InitConds}. We sample $|f_{\vec{k}}|$ on a grid of points in momentum space with the infra-red spacing $1/L$ and the ultra-violet cut-off $\Lambda$; we present data only for the largest system sizes $L$ in which the finite-size effects are minimal\footnote{Some of the data involved sampling the interval with quadratic spacing}. We have also ascertained that the average energy is conserved at least to one part in $10^4$. The dimensional dependence of the late time physics is easily accessible as the spatial dimension $d$ can be varied continuously in these simulations. As we employ a lattice in momentum space, we drop the delta function factors in the continuum theory from this point on. Correlation functions that appear in the rest of the article should rightly be thought of as structure factors.

As advertised, we are looking to use the equilibrium phase diagram to rationalize the behavior of our solutions. To this end,
we quickly review the properties of the finite $T$ equilibrium phases (see Figs.~\ref{fig:phased_1} and \ref{fig:phased_2}).
First, in $d<2$, the system is always disordered for all $T>0$.
Correlations decay to zero exponentially with distance and the inverse of the effective mass, $1/m_{\rm eff}$ plays the role of the correlation length.
$m_{\rm eff}>0$ in this phase.
In $d>2$, there are two phases for $r<r_c$ ($r_c$ is the zero temperature quantum critical point). For $T>T_c(r)$, the system is disordered. As we just discussed, $m_{\rm eff}>0$ in this phase.  For $T<T_c(r)$, the system is ordered. The magnetization $\langle\vec{ \Phi} \rangle$ is non-zero and the two-point function $\langle\vec{\Phi}(\vec{x})\cdot\vec{\Phi}(0)\rangle$ decays as a power law $1/|\vec{x}|^{d-2}$ to the positive constant $|\langle \vec{\Phi} \rangle|^2$. A numerically useful property of such a two-point function is that the volume under the curve on long length scales $|\vec{r}|$ scales as $\sim  |\langle \vec{\Phi} \rangle|^2 |\vec{r}|^{d}$. The effective mass squared is zero in this phase and is, physically, the mass of the $(N-1)$ Goldstone modes. At the critical temperature $T_c$, $m_{\rm eff}^2=0$ and the two-point function decays to zero as $1/|\vec{x}|^{d-2}$.

\subsubsection{Results}

Our first result is the one that we alluded to in Sec.~\ref{Sec:Ergodicity}: the late time effective mass squared tends to a non-negative
constant,  $m_f^2 \ge 0$. This is observed in all our numerical solutions. Certainly, $m_f^2 < 0$ is ruled out on grounds
of stability.
Heuristically, the result follows from the averaging over many momentum modes which oscillate at different frequencies. Once the effective mass settles, the density matrix in the eigenbasis of $H$ dephases and is effectively diagonal.

The limiting behavior of the effective mass divides into two classes with very different physical content.

In the first case, the effective mass squared tends to a positive constant ($m_f^2>0$). It is easily checked that this implies a finite correlation length at late times. The late time behavior is then qualitatively the same as that of the equilibrium disordered state. We emphasize that we are not implying thermalization: $m_f^2$ is not a function of the excess energy density and the angular momenta $\langle L_{\vec{k}}^z\rangle =0$ alone. It depends on many properties of the initial state. Similarly, the form of the late time correlations are not given by the appropriate GGE.

In the second case, $m_{\rm eff}^2$ tends to zero as $t\rightarrow \infty$ and the correlation length diverges in the same
limit. We shall refer to this behavior as coarsening. This case subsumes two cases---that of strict coarsening and that of coarsening to a critical state. By strict coarsening, we mean the analog in our system of the process in an ergodic system
in which the system is quenched from the disordered to the ordered phase and the symmetry is only broken locally \cite{Bray:1994aa}.
The domains of broken symmetry grow with time; their characteristic size at late times, $l_{co}(t)$, grows as $t^{1/z_d}$, where $z_d$ is a dynamic exponent. Within each bubble, the two point function heals to its equilibrium value $|\langle \vec{\Phi} \rangle|^2$ on a length scale $\xi \ll l_{co}(t)$.
On the longer length scale $l_{co}(t)$, it decays to zero. In contrast, by coarsening to a critical state, we mean the analog of the the system approaching a critical state with no length scale except $l_{co}(t)$ and hence no domains exhibiting equilibrium magnetization. In either case, the system is self-similar on the scale $l_{co}(t)$ and a dynamical scaling theory emerges in the limit $t\rightarrow \infty, |\vec{k}|\rightarrow 0$ holding $|\vec{k}|^{z_d} t$ fixed. In this limit, the equal time structure factor has the scaling form:
\begin{align}
\langle \Phi_{\vec{k}}(t)  \Phi^\dagger_{\vec{k}}(t) \rangle \sim \frac{1}{|\vec{k}|^\delta} \mathcal{G}(|\vec{k}|^{z_d} t), \label{Eq:CoarseningScaling}
\end{align}
where $\mathcal{G}$ is the scaling function. When the system is strictly coarsening, the volume under the two-point function should grow as $l_{co}(t)^d$. That is, the structure factor at zero momentum should as $l_{co}(t)^d$. Thus, $\delta = d$ for strict coarsening. On the other hand, when the system is coarsening to a critical state, the growth of the volume under the two-point function with time is slower. Thus, $\delta<d$.

To numerically confirm that the system is coarsening, we will ask that the structure factor at late times have the scaling form predicted by Eq.~\eqref{Eq:CoarseningScaling}. We will then use the value of $\delta$ to differentiate between the two possibilities. Let us now turn to the amplitude and spatial dimension dependence of the late time behavior of our sudden quenches.

\begin{figure}
\begin{center}
\includegraphics{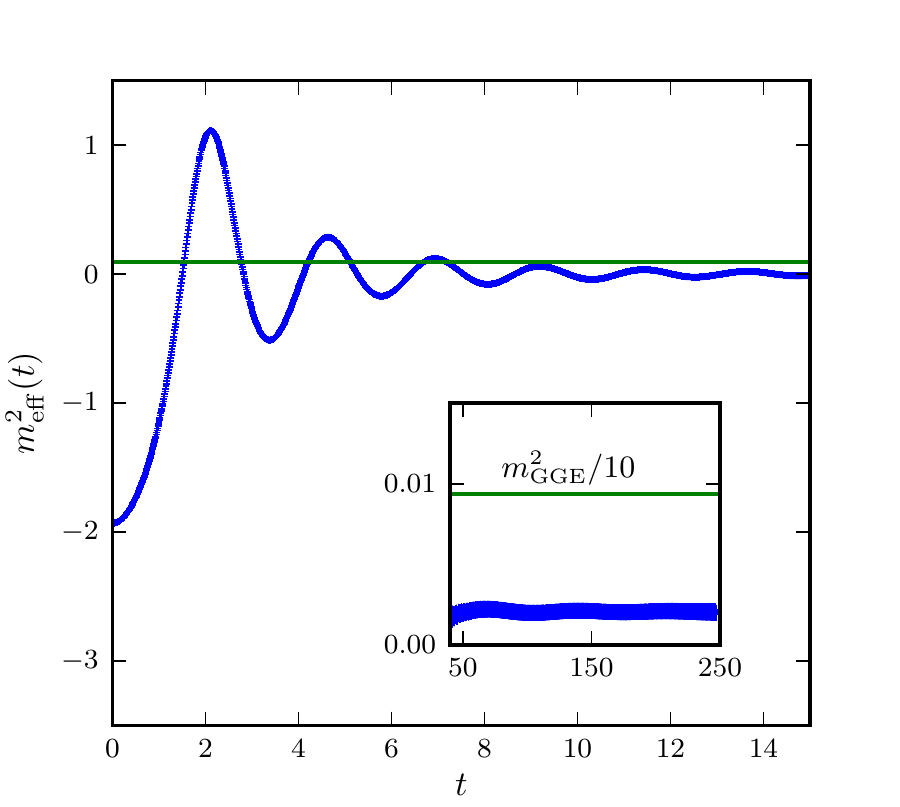}
\caption{Plot of $m_{\rm eff}^2(t)$ vs $t$ for a sudden quench from the disordered phase to the ordered phase in $d=1+\epsilon$. The green line shows $m_{GGE}^2$, the effective mass squared predicted by the GGE. Inset shows late time. System parameters: $\epsilon=0.1$, $L=1200$, $\Lambda=\pi$, $r_0=r_c+2$, $r=r_c-1$, $\lambda=1$.  }
\label{Fig:fig-d1-1-meff2}
\end{center}
\end{figure}

\begin{figure}
\begin{center}
\includegraphics{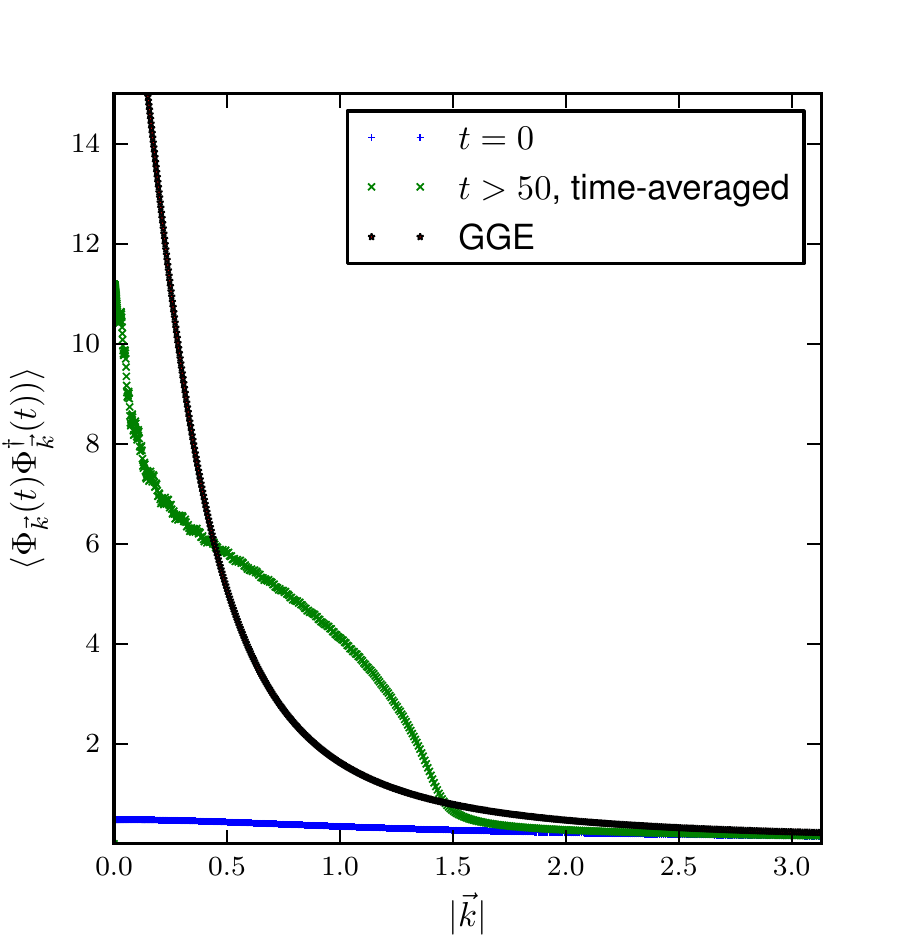}
\caption{Plots of $\langle \Phi_{\vec{k}}(t) \Phi^\dagger_{\vec{k}}(t)\rangle$ vs $|\vec{k}|$ at different times for a sudden quench from the disordered phase to the ordered phase in $d=1+\epsilon$. System parameters: $\epsilon=0.1$, $L=1200$, $\Lambda=\pi$, $r_0=r_c+2$, $r=r_c-1$, $\lambda=1$. }
\label{Fig:StrucFacd1.1}
\end{center}
\end{figure}

Consider first sudden quenches in $d\leq 2$. Relatively small system sizes are sufficient to determine the late-time behavior near $d=1$ rather than at $d=2$. We therefore work at $d=1+\epsilon$. Fig.~\ref{Fig:fig-d1-1-meff2} shows $m_{\rm eff}^2(t)$ for $d=1.1$; the inset contains the longest time behavior. Observe that $m_{\rm eff}^2$ can be negative in the course of the evolution. This is not alarming because it is not the steady state behavior and does not imply instantaneous imaginary correlation lengths. The negative values of $m_{\rm eff}^2$ merely indicates that $\langle \Phi^2(t) \rangle<(-r)$. The mode functions then grow exponentially, causing $m_{\rm eff}^2$ to become positive. This explains the initial oscillatory behavior of $m_{\rm eff}^2$. However, at late times, $m_{\rm eff}^2$ settles to the positive value $m_f^2$ (see inset). The green line in both plots is the effective mass squared predicted by the GGE, $m^2_{GEE}$. The system definitely does not relax to the GGE as $m^2_{GGE}$ is almost two orders of magnitude larger than $m_f^2$. The equal time structure factor is plotted in Fig.~\ref{Fig:StrucFacd1.1} at different times. By the Riemann-Lebesgue lemma, only the time-averaged structure factor contributes to the two-point function in real-space. Thus, the green curve is sufficient to understand the behavior of correlations in real-space. All three structure factors lead to exponential correlations in real-space.

The positivity of $m_{f}^2$ is not specific to our choice of initial conditions. Recall that stability arguments dictate that $m_f^2 \geq 0$ if the effective mass goes to a constant as $t\rightarrow \infty$. We now show analytically that $m_f^2=0$ is physically impossible in $d\leq 2$. To this end,
suppose $m_{f}^2=0$. Then, the solution for $f_{\vec{k}}(t)$:
\begin{align}
f_{\vec{k}} &= A_{\vec{k}} \cos(|\vec{k}| t) + \frac{B_{\vec{k}}}{|\vec{k}|} \sin(|\vec{k}| t),
\end{align}
where $A_{\vec{k}}$ and $B_{\vec{k}}$ are complex valued functions that depend on the initial conditions. Whatever the detailed form, their amplitude must be finite and non-zero and $Im[A_{\vec{k}} B^*_{\vec{k}}]=1$ (from Eq.~\eqref{Eq:Wronskian}). The above solution has to be consistent with $m_{f}^2$ determined through Eq.~\eqref{Eq:meff2inf} (or Eq.~\eqref{Eq:Meff2Rho}). For $d< 2$, this is impossible as the RHS is different from zero by an amount divergent in the long time limit:
\begin{align}
r+\frac{\lambda}{2} \int \frac{d^dk}{(2\pi)^d} |f_{\vec{k}}|^2 \mathcal{N}_{\vec{k}} \sim t^{2-d}.
\end{align}
Therefore, $m_f^2>0$ for $d<2$ and the late time state is disordered. The argument at $d=2$ is more delicate; it involves showing that the RHS is different from zero by a finite quantity.

\begin{figure}
\begin{center}
\includegraphics{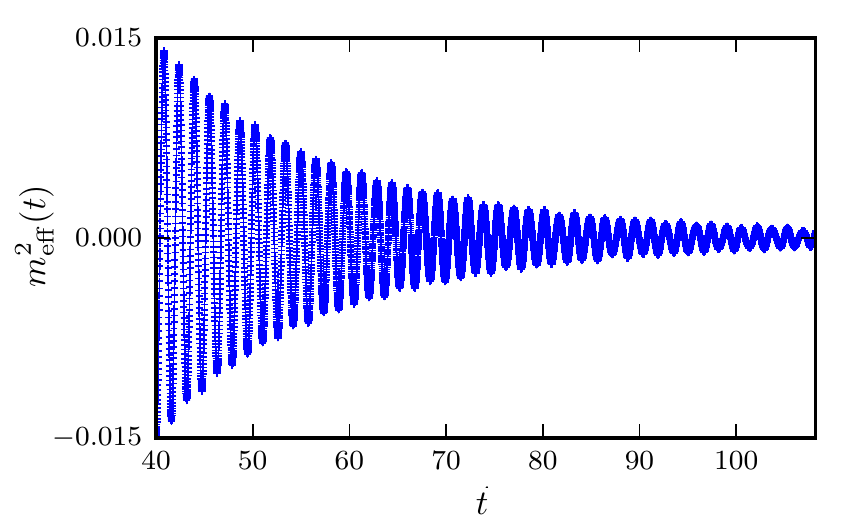}
\caption{Plot of $m_{\rm eff}^2(t)$ vs $t$ for a sudden quench of small amplitude (see text for definition) from the disordered phase to the ordered phase in $d=3$. The ratio of the injected excess energy density to the critical energy density is approximately $0.7$. System parameters: $L=700$, $\Lambda=\pi$, $r_0=r_c+2$, $r=r_c-4$, $\lambda=1/2$. }
\label{Fig:Meff2d3}
\end{center}
\end{figure}

The above proof suggests that for $d>2$ different quenches {\it can} lead to a vanishing $m_{f}^2$ and hence coarsening.
Indeed, our numerical results for $d>2$ confirm this expectation and show two kinds of late time behaviors.
Shown in Fig.~\ref{Fig:Meff2d3} is the plot of $m_{\rm eff}^2(t)$ at late times for a sudden quench of ``small amplitude" to the ordered phase in $d=3$.
We define a ``small amplitude" quench to be a sudden quench in which the injected energy density is smaller than that in the critical ensemble at $r$.
At late times, the effective mass is seen to oscillate about zero with a decaying amplitude. Thus, $m_f^2$ is indeed zero.

\begin{figure}
\begin{center}
\includegraphics{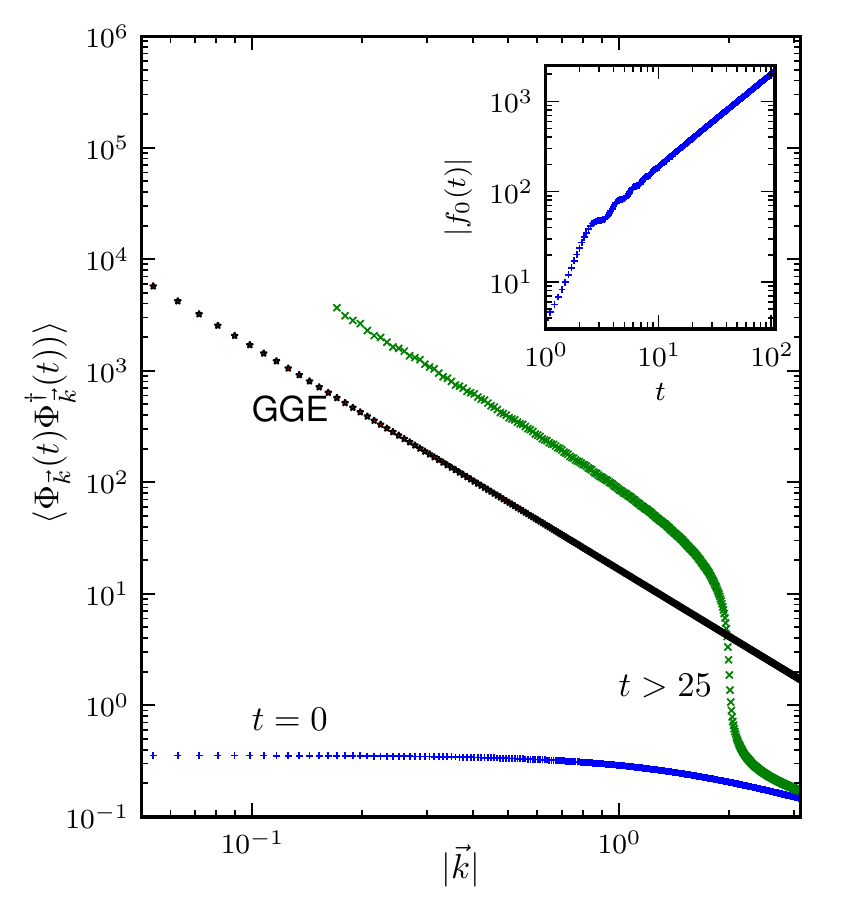}
\caption{Plots of $\langle \Phi_{\vec{k}}(t) \Phi^\dagger_{\vec{k}}(t)\rangle$ vs $|\vec{k}|$ at different times for a quench of small amplitude from the disordered phase to the ordered phase in $d=3$. The green curve is the averaged structure factor for $t>25$ (see text). Inset: The amplitude of the mode function at $|\vec{k}|=0$, $|f_0(t)|$, vs $t$. See the label of Fig.~\ref{Fig:Meff2d3} for system parameters.}
\label{Fig:StrucFacd4}
\end{center}
\end{figure}

In Fig.~\ref{Fig:StrucFacd4}, we plot various structure factors: 1) at $t=0$, 2) the averaged structure factor at late times and 3) the connected structure factor predicted by the GGE. We compute the averaged structure factor only for the wave numbers $|\vec{k}|\gtrsim 0.1$. For $t>25$, their amplitudes, $|f_{\vec{k}}|$, oscillate about the green curve. Again, by the Riemann-Lebesgue lemma, the green curve is sufficient to compute the late time two-point function in real-space at distances $\lesssim 2\pi/0.1$ in lattice units. Observe that despite the lack of relaxation to the GGE, the tail of the green curve falls off as $1/|\vec{k}|^2$, just like the curve computed from the GGE (black). In the inset in Fig.~\ref{Fig:StrucFacd4}, we plot $|f_0(t)|$ as a function of $t$.  At early times, $|f_0(t)|$ is order one as in the initial state. At late times, it grows linearly with $t$. This implies that the volume under the two point function $|f_0|^2$ grows as $t^2$. This is consistent with the scenario of coarsening to a non-equilibrium critical state, as we will show now.

\begin{figure}
\begin{center}
\includegraphics{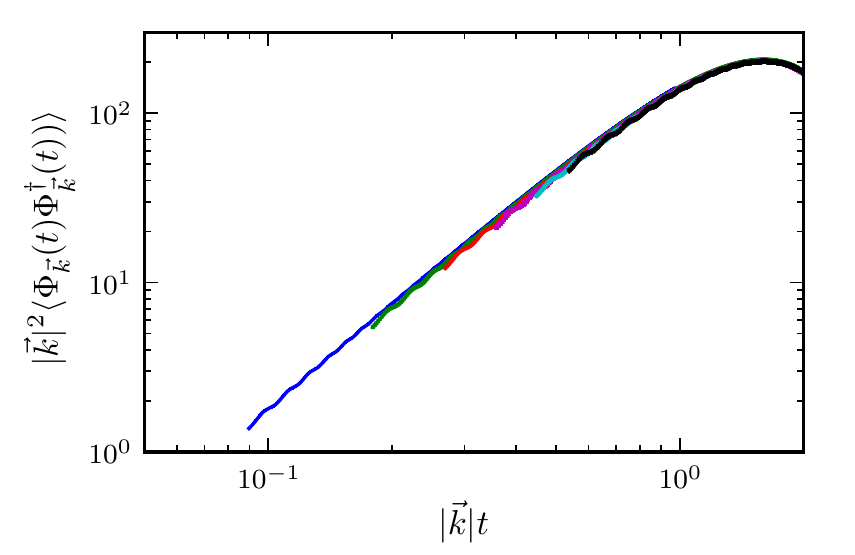}
\caption{The structure factor multiplied by $|\vec{k}|^2$ vs $|\vec{k}|t$ for the quench of small amplitude from the disordered phase to the ordered phase in $d=3$. The six  curves that exhibit the scaling collapse predicted by Eq.~\eqref{Eq:CoarseningScaling} are at the six smallest values of $|\vec{k}|$, $|\vec{k}|=2\pi m/L$, where $m=1,\ldots 6$ and $L=700$. We conclude that $z_d=1, \delta=2$.}
\label{Fig:ScaledSF}
\end{center}
\end{figure}

We now check if the structure factor has the scaling form predicted in Eq.~\eqref{Eq:CoarseningScaling}. Shown in Fig.~\ref{Fig:ScaledSF} is the plot of the scaling function $\mathcal{G}$ vs $|\vec{k}| t$. The curves at different $|\vec{k}|$ collapse when $z_d=1, \delta=2$. As $\delta<d$, the system is coarsening to a non-equilibrium critical state, as opposed to strict coarsening (in agreement with Ref.~\onlinecite{Sciolla:2012ys}). Physically, when $z_d=1$, the size of a correlated region at time $t$ is set by the horizon. This is the fastest possible growth and is possibly too fast to establish long-range order. An interesting avenue for future work is to investigate the nature of the coarsening process in the $O(N)$ model at infinite $N$ limit in different spatial dimensions. Preliminary results show that the system always coarsens to a non-equilibrium critical state. However, the critical state itself varies with dimension \cite{Chandran:2013oq}. 

On injecting an energy density greater than that in the critical ensemble at $r$, the late time behavior is qualitatively the same in Figs.~\ref{Fig:fig-d1-1-meff2},\ref{Fig:StrucFacd1.1}. Again, as was the case for $d\leq2$, the GGE does not reproduce the late time behavior.

\subsubsection{The step approximation}
\label{Sec:StepApprox}
The dimensional dependence of the late time physics is well captured by the step approximation, first introduced in Ref.~\onlinecite{Sotiriadis:2010ys}. Here, one approximates $m_{\rm eff}^2(t)$ by a step function:
\begin{align}
m_{\rm eff}^2(t) \approx
 \begin{cases}
m_0^2, \qquad t <0 \\
m_{s}^2, \qquad t>0.
\end{cases}
\end{align}
The initial state fixes $m_0^2$. This approximation is quite coarse in that it ignores the intricate early-time behavior of $m_{\rm eff}^2(t)$ visible in Fig.~\ref{Fig:fig-d1-1-meff2}. However, it builds in the late-time constancy of $m_{\rm eff}^2(t)$, a key feature of the infinite $N$ dynamics. Within this approximation, the equations of motion can be solved analytically as the dynamical problem is equivalent to a sudden quench in a free field theory. $m_s^2$ is a free parameter in this solution. It is fixed by requiring that the self-consistency relation in Eq.~\eqref{Eq:Meff2Rho} hold at late times. For further details of the method, see Ref.~\onlinecite{Sotiriadis:2010ys}.

Although $m_s$ is numerically \emph{not} equal to the actual late time effective mass, $m_f$, the two share qualitative features. Specifically, the dimensional dependence of $m_f$ discussed in the previous subsection is reflected by $m_s$. First, we find that for $d\leq2$, $m_s^2$ is always positive for any $m_0$. Thus, the final state is always disordered. Second, for $d>2$, the value of $m_s^2$ depends on $m_0^2$. For deep quenches or ``large" $m_0$, $m_s>0$ and the final state is disordered. On decreasing the amplitude of the quench, the value of $m_s^2$ decreases, until $m_s^2=0$. After this point, we find no solutions to the self-consistency relation and the approximation breaks down. This breakdown is indicative of the new physics of coarsening at late times. Note though that coarsening cannot be captured within this approximation as the precise way in which $m_{\rm eff}^2(t)$ approaches zero is important.

To summarize, the step approximation leads to a disordered state in $d\leq 2$ for any quench and in $d>2$ for a deep quench. For shallow quenches in $d>2$, the approximation breaks down, indicating new late-time behavior.

\section{Scaling and Quenches to the critical point}
\label{Sec:ToCriticalPoint}

Thus far we have focused on the dimensional dependence of the late time physics upon quenching to the ordered phase. Now we turn to the question of scaling for such quenches due to proximity to the critical point
separating the ordered and disordered phases. This aspect of the physics
has an interesting dimensional dependence of its own which is the
non-equilibrium analog of the variation of equilibrium critical behavior
with dimension. We will specifically be interested in the reversion to
Gaussian critical behavior in $d \ge 3$. We will focus almost entirely
on quenches to the critical point and note at the end the generalization
to quenches into the ordered phase.

As before, we prepare the system in the ground state in the disordered phase at some $r_0>r_c$. The initial conditions are once again as detailed in
Eq.~(\ref{Eq:InitConds}). At $t=0$, we quench the system to the critical point.

A first observation is that in any dimension the late time state is disordered, i.e. $m_{f}^2 >0$ for an initial non-zero $m_{0}$. This is observed in our numerical solutions but it can also be argued, on analytic grounds paralleling
our demonstration that coarsening is impossible upon quenching into the ordered phase. For $d < 2$, our previous argument already suffices and for $d>2$, one
can extend it via a consideration of a larger set of momenta\cite{Chandran:2013oq} at
the critical coupling. We note that this result is again consistent with reasoning based on the equilibrium phase diagram---injecting a non-zero energy density at
 the critical coupling should lead to a finite correlation length. However, this result hides an important difference
between $d<3$ and $d\geq 3$ to which we now turn. This difference is most
sharply visible if we consider a scaling limit for sudden quenches, formulated
in Ref.~\onlinecite{Chandran:2012qa} which makes their universal physics manifest.

Let us first consider $1<d<3$ where the critical point is interacting even in the infinite $N$ theory. When the initial state is in the vicinity of this critical point, its correlation length,  $1/m_0^{1/z}$, diverges as ($r_0 - r_c)^{-\nu}$, where the critical exponents are $\nu = 1/(d-1), z=1$ at infinite $N$. On the length/time scale $1/m_0$, the equilibrium physics is universal. On the same scale, the dynamics following a sudden quench is also universal in an appropriate scaling limit. This is the limit of $r_{0}\rightarrow r_c$ when lengths/times are measured in units of the initial correlation length/time, $1/m_{0}$. In this limit, the one and two point functions of an operator $O$ have the scaling forms:
\begin{align}
\langle O(\vec{x}, t) \rangle_{m_{0}} &\sim m_0^\Delta \,\mathcal{G}_{\scriptscriptstyle O}( t \,m_{0}) \\
\langle O(\vec{x}, t) O(0, 0) \rangle_{m_{0}} &\sim m_0^{2\Delta}\,\mathcal{G}_{\scriptscriptstyle OO} ( |\vec{x}| \,m_{0},t \,m_{0}), \label{Eq:2ptscalingform}
\end{align}
where $\Delta$ is the scaling dimension of the operator $O$ and we have used the translational and rotational invariance of the system. The out-of-equilibrium physics in this limit is conjectured to be universal because the quench is very shallow -- only the universal low-energy long-wavelength part of the energy spectrum at the critical point is excited for $t>0$. Those readers familiar with the Kibble-Zurek mechanism will notice that the above scaling limit is very similar to the one in slow ramps. The correlation length and time in the initial state play the roles of the Kibble-Zurek length and time.

The previous computations of universal correlation functions in $d=1$ are in the scaling limit described above. For example, the two-point correlation functions computed by Calabrese and Cardy\cite{Calabrese:2006aa, Calabrese:2007aa} using boundary conformal field theory have the scaling form in Eq.~\eqref{Eq:2ptscalingform}. Finally, the formalism above can be easily generalized to the cases when (a) the dynamic exponent $z\neq 1$, (b) when the system size is finite and (c) when the quench is not exactly to the critical point.

\begin{figure}
\begin{center}
\includegraphics{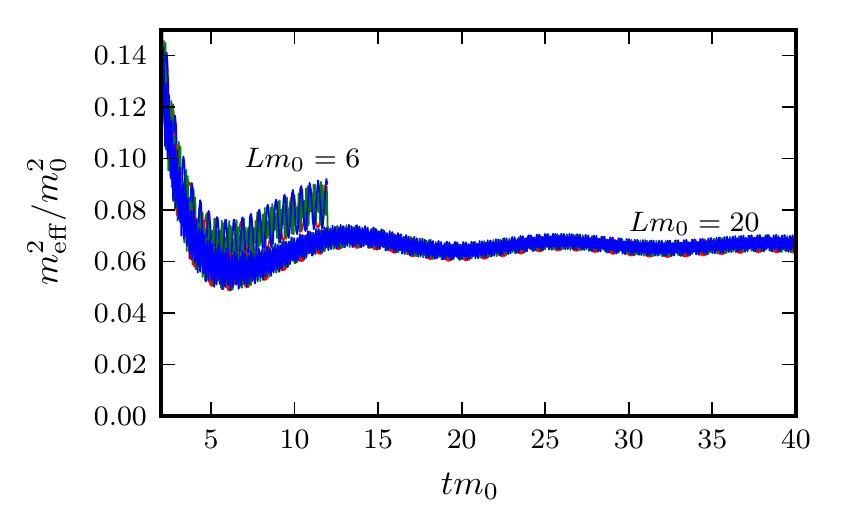}
\caption{Plots of $m_{\rm eff}^2/m_0^2$ vs $t m_0$ at different values of $L m_0$ in d=2. At each $Lm_0$, we show the behavior for three (small) values of $m_0$. The scaling collapse provides strong evidence for Eq.~\eqref{Eq:msqscalingform}. System parameters: $\Lambda=1$, $r=r_c$, $\lambda=3$.}
\label{Fig:meff2scalingcollapse}
\end{center}
\end{figure}

\begin{figure}
\begin{center}
\includegraphics[width=8.3cm]{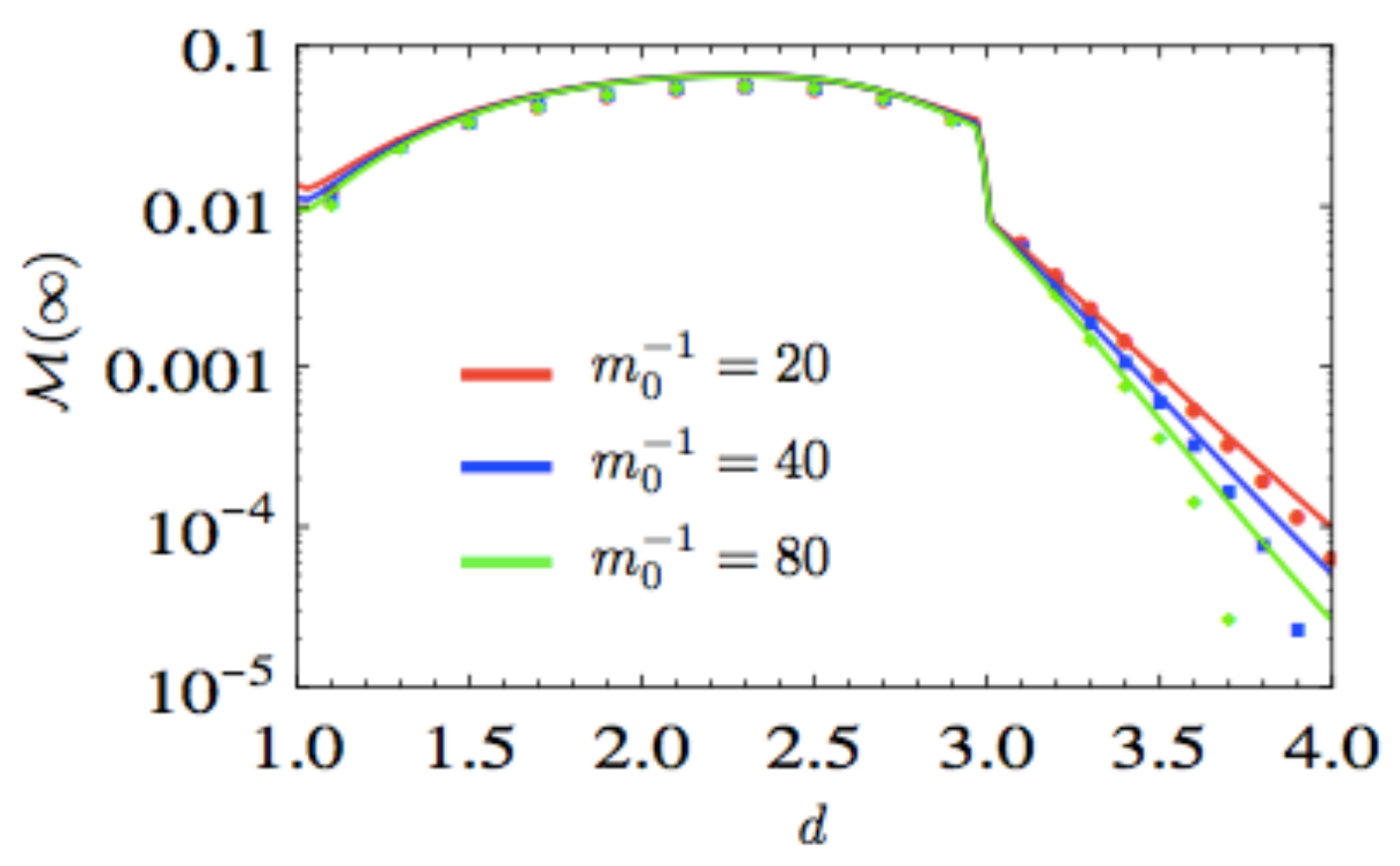}
\caption{$\mathcal{M}(\infty)$ vs spatial dimension $d$ computed for $m_0^{-1} = 20, 40, 80$ with $Lm_0=10$. The solid line is $\mathcal{M}(\infty)$ obtained in the step approximation (Sec.~\ref{Sec:StepApprox}) at the same system system sizes and initial conditions. System parameters: $\Lambda=1$, $r=r_c$, $\lambda$ is $d$-dependent and chosen to ensure that the initial state is in the equilibrium scaling region.}
\label{Fig:Cd_SC}
\end{center}
\end{figure}

To test the scaling hypothesis, we consider the one-point function, $\langle |\vec{\Phi}|^2(t) \rangle_{m_0}/N$. Equivalently, consider the effective mass squared $m_{\rm eff}^2$. Accounting for the finite system size $L$ in the numerics, the scaling form of $m_{\rm eff}^2$ is:
\begin{align}
m_{\rm eff}^2(t, L ;m_0)\sim m_0^2 \mathcal{M}(t \, m_0, L\, m_0). \label{Eq:msqscalingform}
\end{align}
We discuss the early and late time behavior of $\mathcal{M}$ below.

By construction in the thermodynamic limit, $\mathcal{M}(0^-)=1$. At $t=0^+$, $m_{\rm eff}^2$ is negative. This reflects the finite correlations in the initial state, unlike the ground state at the critical point. Using the form of $f_{\vec{k}}(0^-)$ in Eq.~\eqref{Eq:Meff2Rho}, it is easy to see that:
\begin{align}
\mathcal{M}(0^+)\sim-m_0^{d-3}.
\end{align}
Thus, $\mathcal{M}(0^+)$ is negative and divergent in the scaling limit.

The full scaling function can be determined numerically. In Fig.~\ref{Fig:meff2scalingcollapse}, we show $\mathcal{M}$ for $L m_0=6,20$ in $d=2$. The data collapses for three different values of $m_0$ at each $L m_0$, providing strong evidence for Eq.~\eqref{Eq:msqscalingform}. The negative divergence of $\mathcal{M}$ near $0^+$ is not shown in order that the late time behavior be visible. Generically, we find that $\mathcal{M}$ tends to a positive constant at late times. Thus, the late time state is indeed disordered.

In Fig.~\ref{Fig:Cd_SC}, $\mathcal{M}(\infty)$ is plotted as a function of dimension for different values of $m_0$ at fixed $L m_0 = 10$. For all $d<3$, the extrapolation to the thermodynamic limit leads to a positive value of $\mathcal{M}$ as $t m_0 \rightarrow \infty$. The solid line is $\mathcal{M}(\infty)$ in the step approximation discussed in Sec.~\ref{Sec:StepApprox}. In agreement with Eq.~\eqref{Eq:msqscalingform}, $\mathcal{M}(\infty)$ is finite in the dynamic scaling limit. The close agreement with the numerical data points appears fortunate rather than principled.

Now consider the case when $d \ge 3$ where the fixed point is Gaussian. If we work directly at the fixed point, we have a free field theory, and
the time-dependent mode functions can be calculated exactly for the sudden quench to the critical point (see, for example Ref.~\onlinecite{Calabrese:2007aa}).
In the language of this article, the computation amounts to ignoring the self-consistency equation for the effective mass and assuming that it is zero at all times.
The result is that the correlation functions at late times have a power-law behavior in real space at long distances (light cone effects are irrelevant to this discussion). Now, this is at odds with the observation of a \emph{finite} correlation length in Fig.~\ref{Fig:Cd_SC} at any \emph{non-zero} $m_0$. This discrepancy is resolved if we take proper account of the irrelevant quartic term which is dangerously irrelevant already for the
equilibrium behavior, i.e.~it cannot be neglected to get proper asymptotic results. 
The same is true for the non-equilibrium dynamics. This is easy to see in the step approximation. The dependence of the late time (finite) correlation length, $\xi_f$, on the initial correlation length $1/m_0$ is given by:
 \begin{align}
\xi_f m_0 \sim m_0^{-(d-3)/2} \textrm{ as } m_0 \rightarrow 0.
\end{align}
In the marginal $d=3$ case, $\xi_f$ is only logarithmically greater than $m_0$.
Thus, if we take the scaling limit keeping distances on the order of  $1/m_0$ fixed as described in Ref.~\onlinecite{Chandran:2012qa}, we will indeed find Gaussian behavior but this will no longer
be a good description of the truly long time asymptotics on the longest time and length scales. As in typical problems with dangerously irrelevant variables, there are
now two divergent scales to contend with---a phenomenon likely to be much more common in the non-equilibrium setting as has been noted already with a different
example involving string-net coarsening in Ref.~\onlinecite{Chandran:2012qf}. Finally, we note that the dangerous irrelevance of interactions is also germane to quenches into the ordered phase where again a scaling limit can be defined as above. Again, we will find that in $d \ge 3$ Gaussian results hold in the scaling regime, which will exhibit an exponential growth of the local order parameter before being cutoff at the parametrically longer scale by the coarsening physics we described previously.

\section{Concluding remarks}
\label{Sec:Conclusion}

In our current understanding, as sketched in this paper, the quantum $O(N=\infty)$ vector model appears intermediate between generic systems that
exhibit thermalization starting out of equilibrium and integrable systems that do not. It does exhibit stationary behavior at long times following
parameter changes but it does not exhibit thermalization. This behavior is consistent with it appearing to have only half the number of conserved 
quantities appropriate for a fully integrable system. Interestingly, the GGE constructed from the known conserved quantities does {\it not} describe
the late time stationary states which is inconsistent with the GGE conjecture \cite{Rigol:2007cr}. A definitive resolution to
the question of whether the GGE conjecture is false or whether there are additional conserved quantities is desirable. In the absence of the latter
the $O(N=\infty)$ vector model would provide an example of a system exhibiting a large number of emergent conserved quantities at late times.  In such a case, at least the most salient properties of the system might be captured in terms of a modified GGE which includes chemical potentials for the emergent conserved quantities.
\\
Interestingly, the late time states of the $O(N=\infty)$ vector model following quenches exhibit ``knowledge'' of its equilibrium phase diagram. 
They exhibit coarsening if and only if the model exhibits a finite temperature ordered phase, which in turn depends on the dimensionality of the system. 
However, this coarsening process appears to be to a state that is critical. That is, it does not show any signatures of developing long range order 
on length scales smaller than the coarsening length scale. 
Elucidating the nature of this critical coarsening and its precise dependence on initial conditions is a fit subject for future work.
\\
We note that the temporal structure of relaxation appears to be quite complicated at large but finite $N$. The infinite $N$ theory already exhibits finite
times scales for the appearance of stationary states in the absence of coarsening and algebraic relaxation in the presence of coarsening. This
behavior is initial state dependent. For large $N$ we expect this behavior to give way to genuine thermalization, at least for $d>1$, on a time scale 
that is parametrically large in $N$. How much of this intricate time dependence survives to, say, $N=3$ is an interesting question.
\\
Finally we note that in $d=3$ the critical fixed point is Gaussian. In its neighborhood, the scaling limit will be described by the Gaussian
theory as we illustrated in this paper for quenches to the critical point. This is potentially amenable to experimental work in cold atomic
systems where the $N=2$ case is the critical theory of the Mott insulator to superfluid transition with particle hole symmetry.

\begin{acknowledgments}
The authors would like to thank V. Oganesyan, F. Essler, M. Rigol and N. Andrei for stimulating discussions. We are especially grateful to David Huse for many discussions, reading the draft and pointing out that our data was not, in fact, agnostic between different coarsening scenarios. AC and SLS would like to thank the KITP for
hospitality during the Fall 2012 program ``Quantum Dynamics in Far from Equilibrium Thermally Isolated Systems''. 
This work was supported in part by NSF grants DMR 1006608, PHY-1005429 and NSF PHY11-25915(AC,SLS) and in part by the Department of Energy under Grant 
No. DE-FG02-91ER40671 (SSG).

\end{acknowledgments}

\appendix
\section{Lack of relaxation to the GGE}
\label{App:WhynoGGE}
In this appendix, we show that the correlations in the late time state of the $O(N)$ model are not reproduced by the appropriate GGE Eq.~\eqref{Eq:GGEON}. Define a new set of variables linearly related to the mode functions $f_{\vec{k}}$ in Eq.~\eqref{Eq:Fixedbasis}:
\begin{align}
g_{\vec{k}}\equiv f_{\vec{k}}\sqrt{\mathcal{N}_{\vec{k}}}.
\end{align}
We interpret $g_{\vec{k}}$ as the complex coordinate of a 2d classical particle labelled by $\vec{k}>0$. Next, define the momentum of each particle to be:
\begin{align}
\pi = \dot{g}^*.
\end{align}
When absent, the momentum subscript is implied. Then, the condition in Eq.~\eqref{Eq:Wronskian} implies that the angular momentum of the particle at $\vec{k}>0$ is given by:
\begin{align}
 l^z \equiv -Im[g(t) \pi(t)] = -\mathcal{N}.
 \end{align}
This classical angular momentum should not be confused with the conserved angular momentum $L^z$ in Eq.~\eqref{Eq:QuantumLzDef}. They are completely unrelated. The above equation implies that $l^z$ is a constant of the motion for the classical system and is set by the value of $\mathcal{N}$. The dynamics of the classical particles is governed by the Hamiltonian:
\begin{align}
H_{cl} =\int_{ \vec{k}>0} |\pi|^2+ (|\vec{k}|^2 + r) |g|^2+ \frac{\lambda}{4} \left(\int_{ \vec{q}>0}|g|^2\right)^2. \label{Eq:HclassicalFT}
\end{align}
It is easily checked that the equations of motion derived from $H_{cl}$ reproduce Eq.~(\ref{Eq:ModeFunction},\ref{Eq:meff2inf}). On solving for the dynamics of the classical particles generated by $H_{cl}$ with fixed angular momentum $l^z$ and the initial conditions derived from Eq.~\eqref{Eq:cnumberst0}, we can compute any observable in the $O(N)$ model through Eq.~\eqref{Eq:Fixedbasis}. This classical interpretation is useful because it unambiguously identifies the role of the c-number $\mathcal{N}$ in the solution. The radial coordinate of the classical particle, $|g(t)|$, depends on the conserved angular momentum $l^z$ at late times. As the equal time structure factor $S(\vec{k},t)$ is given by:
\begin{align}
S(\vec{k},t) &=\frac{|g_{\vec{k}}(t)|^2}{2} \\
\textrm{where } \langle \Phi_{\vec{k}} \Phi^\dagger_{\vec{k'}} \rangle &=  S(\vec{k},t) \, (2\pi)^d \delta^d(\vec{k}-\vec{k}'),
\end{align}
the time-averaged structure factor at late times must depend on $l^z_{\vec{k}}$ as well. The structure factor in the GGE however is independent of this quantity. Thus, the two do not agree and the dynamics in the $O(N)$ model does not thermalize in the generalized sense.

\bibliography{QuantumLargeN}
\end{document}